\documentclass[superscriptaddress,amsmath,amssymb,aps,prl,twocolumn]{revtex4}
\usepackage{graphicx}
\usepackage{amsmath}
\usepackage{amssymb}
\usepackage{braket}
\usepackage{dcolumn}
\usepackage[colorlinks=true,linkcolor=blue,citecolor=blue,urlcolor=blue]{hyperref}

\begin{document}

\title{Krylov Complexity in Open Quantum Systems}
\author{Chang Liu}
\thanks{They contribute equally to this work. }
\affiliation{Institute for Advanced Study, Tsinghua University, Beijing, 100084, China}
\author{Haifeng Tang}
\thanks{They contribute equally to this work. }
\affiliation{Institute for Advanced Study, Tsinghua University, Beijing, 100084, China}
\affiliation{Department of Physics, Tsinghua University, Beijing, 100084, China}
\author{Hui Zhai}
\email{hzhai@tsinghua.edu.cn}
\affiliation{Institute for Advanced Study, Tsinghua University, Beijing, 100084, China}
\date{\today}

\begin{abstract}

Krylov complexity is a novel measure of operator complexity that exhibits universal behavior and bounds a large class of other measures. In this letter, we generalize Krylov complexity from a closed system to an open system coupled to a Markovian bath, where Lindbladian evolution replaces Hamiltonian evolution. We show that Krylov complexity in open systems can be mapped to a non-hermitian tight-binding model in a half-infinite chain. We discuss the properties of the non-hermitian terms and show that the strengths of the non-hermitian terms increase linearly with the increase of the Krylov basis index $n$. Such a non-hermitian tight-binding model can exhibit localized edge modes that determine the long-time behavior of Krylov complexity. Hence, the growth of Krylov complexity is suppressed by dissipation, and at long-time, Krylov complexity saturates at a finite value much smaller than that of a closed system with the same Hamitonian. Our conclusions are supported by numerical results on several models, such as the Sachdev-Ye-Kitaev model and the interacting fermion model. Our work provides insights for discussing complexity, chaos, and holography for open quantum systems. 

\end{abstract}

\maketitle

Operator complexity describes how an operator becomes increasingly complicated under the Heisenberg time evolution. The concept of operator complexity has emerged as a novel tool in studying quantum matters \cite{B. Yoshida, R.C. Myers, A. Streicher, R.-Q. Yang, S. Sharma, K.-Y. Kim, A. Streicher-2, A.Lucas, O. Parrikar, O. Parrikar-2}. It can characterize the chaotic behavior and integrability of a quantum many-body Hamiltonian, and it is correlated with the dynamics of quantum information processes. Through holography, it also becomes a new entity to study black hole physics. A mathematically rigorous definition of operator complexity depends on the choice of a pre-defined basis. Previously, various measures of operator complexity have been proposed and studied in different contexts \cite{B. Yoshida, R.C. Myers, A. Streicher, R.-Q. Yang, S. Sharma, K.-Y. Kim, A. Streicher-2, A.Lucas, O. Parrikar, O. Parrikar-2}. 

Recently, the Krylov recursion method has been applied to investigate operator complexity \cite{Altman}. It is proposed that the operator complexity in the Krylov basis, called Krylov complexity, exhibits universal behaviors and can bound a large class of other measures \cite{Altman}. Thanks to its advantages, Krylov complexity has attracted considerable attention from various communities \cite{R. Sinha, A. Gorsky, X. Cao, A. Dymarsky, A. Gorsky-2, M. Sasieta, J. Simon, Z. Y. Xian, J. Sonner, A. Lucas,  M. Smolkin, J. D. Noh, C. J. Lin, D Shouvik, D. Patramanis, D. Patramanis-2, J. Sully, D. Rosa, A. del Campo, J. Sonner-2, T. Pathak, Q. Wu, J. Gemmer, B. Swingle, S Choudhury, A. Roy, S Liu, Y. Yang, S Pawar, Z. Y. Fan, Z. Y. Fan-2,  J. Sonner-3}. Nevertheless, research on Krylov complexity has so far been limited to closed systems. In this letter, we generalize Krylov complexity from a closed system to an open system. In a closed system, operator growth is governed by Hamiltonian, while for an open system coupled to a Markovian bath, operator growth is governed by Lindbladian. Here we will discuss how this change from Hamiltonian to Lindbladian affects the behavior of Krylov complexity.

\begin{figure}
\begin{center}
\includegraphics[width=0.45\textwidth]{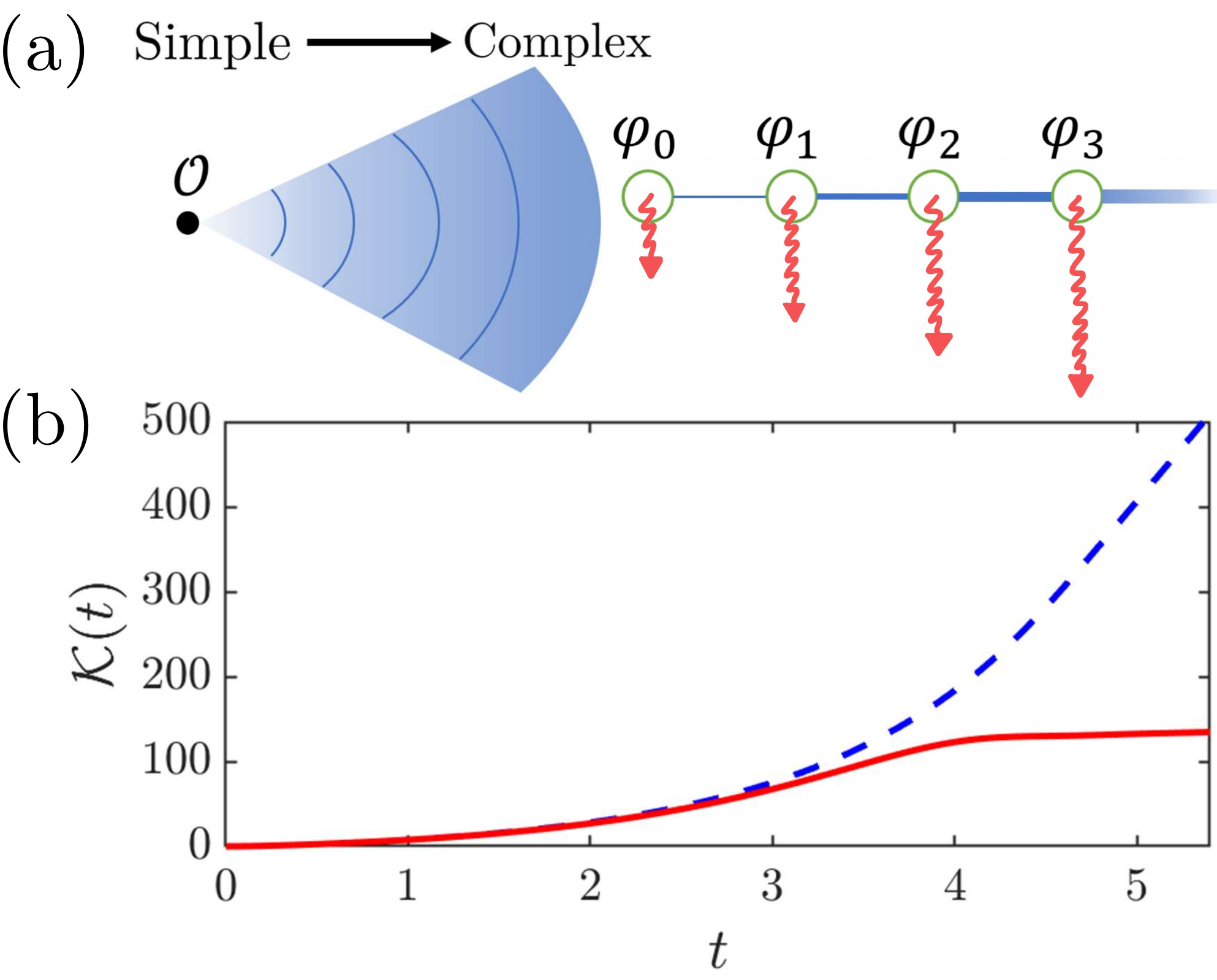}
\end{center}
\caption{(a) Schematic the mapping between Krylov complexity in open systems and a non-hermitian tight-binding model in a half-infinite chain. (b) Krylov complexity $\mathcal{K}(t)$ in open systems (red solid line), compared with $\mathcal{K}(t)$ in a closed system with the same Hamiltonian (blue dashed line). Krylov complexity is suppressed by dissipation.   }
\label{results}
\end{figure}

\textit{Review of Krylov Complexity.} Before starting the generalization, let us briefly review Krylov complexity in a Hamiltonian system \cite{Altman}. First of all, for a system with its Hilbert space spanned by $\{\ket{i}\}$, an operator $\hat{X}=\sum_{ij}X_{ij}\ket{i}\bra{j}$ can be mapped to a state in the double space, denoted by $\ket{\hat{X}}=\sum_{ij}X_{ij}\ket{i}\otimes\ket{j}$.  We introduce a super-operator $\hat{\mathcal{L}}$ acting on an operator $\hat{X}$ as 
$\hat{\mathcal{L}}\hat{X}=[\hat{H},\hat{X}]$, and $\ket{\hat{\mathcal{L}}\hat{X}}$ is the state corresponding to the operator $\hat{\mathcal{L}}\hat{X}$. Hence, using the Baker-Campbell-Hausdorff formula, the Heisenberg evolution of a reference operator $\hat{O}(t)=e^{i\hat{H}t}\hat{O}e^{-i\hat{H}t}$ can be expressed as expanding the state $\ket{\hat{O}(t)}$ in a set of basis $\ket{\hat{\mathcal{L}}^n\hat{O}}$ as
\begin{equation}
\ket{\hat{O}(t)}=\sum_{n}\frac{(it)^n}{n!}\ket{\hat{\mathcal{L}}^n\hat{O}}. \label{BCH_wf}
\end{equation}
However, this set of basis $\ket{\hat{\mathcal{L}}^n\hat{O}}$ is neither normalized nor orthogonal. Hence, we first need to apply the Gram-Schmidt procedure with the infinite-temperature inner product $\langle O_1|O_2\rangle = \text{Tr}[O_1^\dagger O_2] $ to orthogonalize this set of basis. This results in the Krylov basis $\{\ket{\hat{\mathcal{W}}_n}\}$ as 
\begin{align}
&\ket{\hat{\mathcal{W}}_0}=\frac{1}{b_0}\ket{\hat{O}}, \label{w0} \\
&\ket{\hat{\mathcal{W}}_1}=\frac{1}{b_1}\ket{\hat{\mathcal{L}}\hat{\mathcal{W}_0}}, \label{w1} \\
&\ket{\hat{\mathcal{W}}_n}=\frac{1}{b_n}\left(\ket{\hat{\mathcal{L}}\hat{\mathcal{W}}_{n-1}}-b_{n-1}\ket{\hat{\mathcal{W}}_{n-2}}\right) \   \ \text{for} \   \  n\geqslant 2. \label{wn}
\end{align}
Here $\{b_n\}$ are called the Lanczos coefficients introduced to normalize these states. It is discussed that $b_n$ increases linearly in $n$ for a generic chaotic Hamiltonian \cite{Altman}, until it saturates at large enough $n$ for a finite system \cite{R. Sinha, Z. Y. Xian, J. Sonner, J. Sonner-3}. Therefore, we can expand the state $\ket{\hat{O}(t)}$ under the Krylov basis as 
\begin{equation}
\ket{\hat{O}(t)}=\sum_n \varphi_n(t)\ket{\hat{\mathcal{W}}_n}. \label{expansionW}
\end{equation}
$\varphi_n(t)$ satisfy the following tight-binding model that describes single-particle hopping in a half-infinite chain \cite{Altman}
\begin{equation}
i\partial_t \varphi_n=-b_{n+1}\varphi_{n+1}-b_n\varphi_{n-1}. \label{diff}
\end{equation}
This particle sits at $n=0$ with only $\varphi_0$ being non-zero at $t=0$, and hops away from $n=0$ at finite $t$. Krylov complexity $\mathcal{K}(t)$ is defined as the mean distance measured from $n=0$ as
\begin{equation}
\mathcal{K}(t)=\sum_{n}n|\varphi_n(t)|^2.
\end{equation}

\textit{Summary of Results.} Our generalization of Krylov complexity to open systems keeps using the Krylov basis defined in Eq. \ref{w0}-\ref{wn}. However, the operator dynamics is changed from the Heisenberg evolution to the Lindblad evolution. A Lindbladian contains both the Hamiltonian part and the dissipative part with the dissipation operator $\hat{M}$. The main results are schematically shown in Fig. \ref{results} and summarized as follows. We emphasize that these results are also universal for chaotic Hamiltonians with generic dissipations. 

1) Krylov complexity in an open system can also be mapped to a particle hopping in a half-infinite chain, but described by a non-hermitian tight-binding model as 
\begin{equation}
i\partial_t \varphi_n=-b_{n+1}\varphi_{n+1}-b_n\varphi_{n-1}-i \gamma \sum_{m}d_{nm}\varphi_m, \label{diff-non-hermitian}
\end{equation}
where $\gamma$ represents the dissipation strength. $d_{nm}$ are dominated by their diagnoal terms $d_{nn}$. 

2) For hermitian dissipation operator $\hat{M}$, $d_n$ is always positive. And for a generic chaotic Hamiltonian, $d_n$ grows linearly in $n$ until it saturates at $n>n_\text{s}$. $n_\text{s}$ increases linearly with the increasing of the system size. 

3) When $\gamma>\gamma_\text{c}$, this imaginary part of the spectrum of this non-hermitian tight-binding model exhibits a gap. The wave functions of the modes below the gap are localized at the edge. $\gamma_\text{c}$ decreases toward vanishing when $n_\text{s}$ increases. 

4) The growth of Krylov complexity is suppressed by dissipation. For $\gamma>\gamma_\text{c}$, the localized modes below the gap dominates the long-time evolution of Krylov complexity, therefore, at long-time,  Krylov complexity saturates to a value much smaller the fully scrambled case.     

\begin{figure}[t]
\begin{center}
\includegraphics[width=0.48\textwidth]{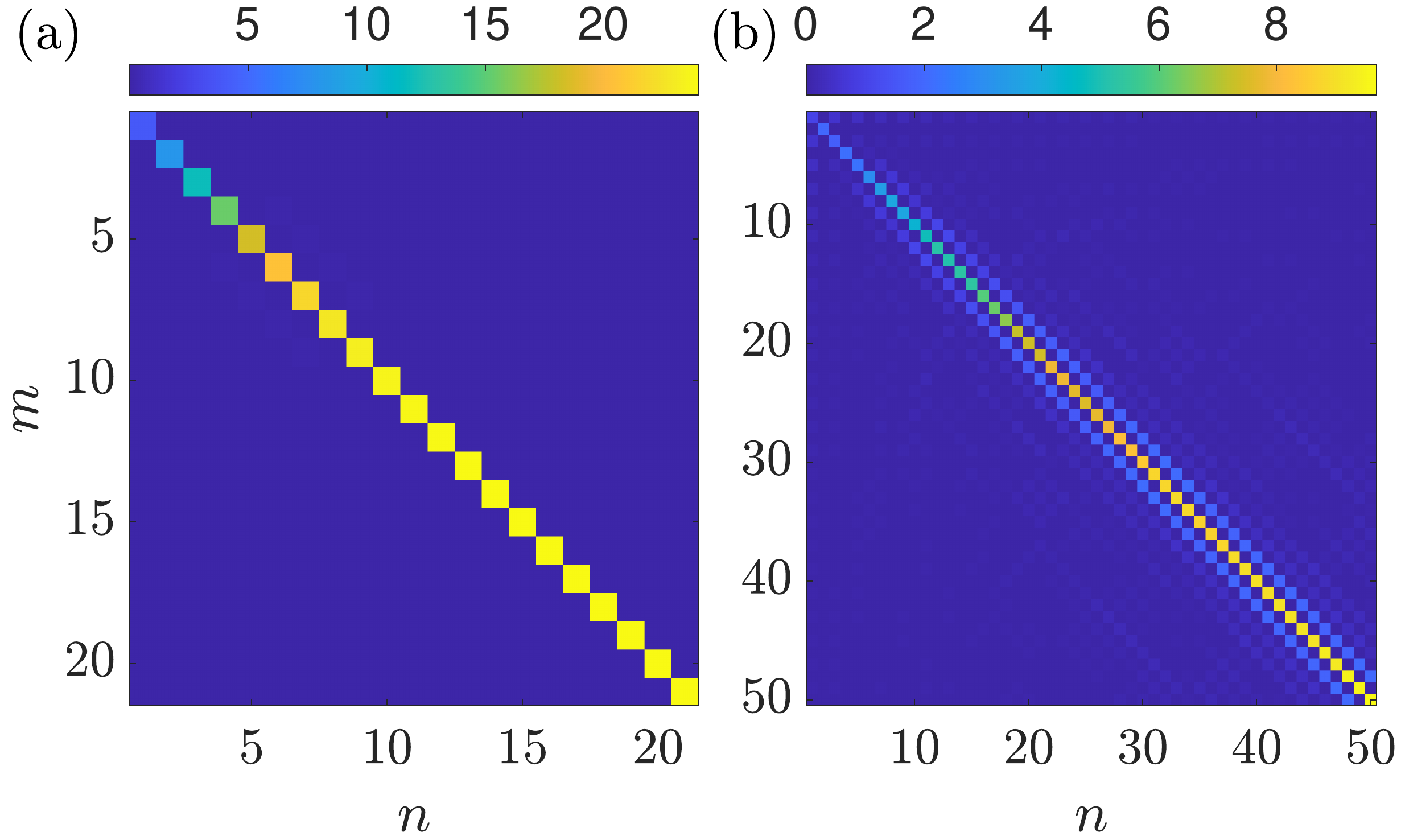}
\end{center}
\caption{Coefficients of the non-hermitian terms $|d_{nm}|$ for the SYK model (a) and for the spinless fermion model (b). For (a), we have chosen $\hat{M}_i=\hat{\psi}_i$ and $\hat{O}=i\hat{\psi}_1\hat{\psi}_2$, and we have set $J=1$. The system contains total number of Majorana fermion $N=24$.  For (b), we have chosen $\hat{M}_i=\hat{c}^\dag_i+\hat{c}_i$ and $\hat{O}=\hat{n}_{1}$, and we have set $J_1=1$, $J_2=0.2$, $V_1=0.6$ and $V_2=0.1$. The model contains total number of sites $N=13$. }
\label{dnm}
\end{figure}

\textit{Non-hermitian Tight-Binding Model.} Now we illustrate these results in detail. First of all, we consider the dynamical equation of an operator $\hat{O}$ under the Lindblad evolution as 
\begin{equation}
\frac{d\hat{O}(t)}{dt}=i[\hat{H},\hat{O}]+\gamma\sum\limits_{i}(\pm 2\hat{M}^\dag_i\hat{O}\hat{M}_i-\{\hat{M}^\dag_i\hat{M}_i,\hat{O}\}), \label{Lindblad_operator}
\end{equation} 
where $\hat{M}_i$ are dissipation operators. Here we note that the minus sign should be taken when both $\hat{O}$ and $\hat{M}_i$ are fermionic operators \cite{F. Schwarz,supple}. This is crucial for following discussion and we present the detailed derivation and explain the origin of this minus sign in the supplementary material \cite{supple}. 
Substituting Eq. \ref{expansionW} into both sides of Eq. \ref{Lindblad_operator}, we arrive at Eq. \ref{diff-non-hermitian} and $d_{nm}$ is given by 
\begin{equation}
d_{nm}=\sum\limits_{i}\text{Tr}[\hat{W}^\dag_n\{\hat{M}^\dag_i\hat{M}_i,\hat{W}_m\}\pm 2\hat{W}^\dag_n \hat{M}^\dag_i \hat{W}_m\hat{M}_i]. \label{dnm}
\end{equation}
Note that the total weight $\mathcal{Z}=\sum_n|\varphi_n|^2$ is conserved in the hermitian case but is not conserved in the non-hermitian case. Nevertheless, we still define the Krylov complexity as the mean distance measured from $n=0$ in the half-infinite chain, and the definition now needs to be modified as 
\begin{equation}
\mathcal{K}(t)=\frac{1}{\mathcal{Z}}\sum_{n}n|\varphi_n(t)|^2.
\end{equation}

\textit{Illustrating Examples.} To illustrate the physics concretely, we consider two representative models. The first model is the Sachdev-Ye-Kitaev (SYK) model \cite{A. Kitaev, S. Sachdev, J. Maldacena}. The Hamiltonian reads 
\begin{equation}
\hat{H}_{\text{S}}=\sum\limits_{i<j<k<l}J_{ijkl}\hat{\psi}_i\hat{\psi}_j\hat{\psi}_k\hat{\psi}_l,
\end{equation}
where $\hat{\psi}_i$ $(i=1,\dots,N)$ denotes $N$ Majorana fermions in the system. $J_{ijkl}$ are independent random Gaussian variables with variances given by $\overline{J^2_{ijkl}}=3!J^2/N^3$. The second model is a one-dimensional lattice model of interacting spinless fermions. The Hamiltonian reads
\begin{align}
\hat{H}_{\text{F}}=&-\sum\limits_{i}\left(J_1\hat{c}^\dag_i\hat{c}_{i+1}+J_2\hat{c}^\dag_i\hat{c}_{i+2}+\text{h.c.}\right) \nonumber\\
&+\sum\limits_{i}\left(V_1\hat{n}_i\hat{n}_{i+1}+V_2\hat{n}_i\hat{n}_{i+2}\right),
\end{align}
where $J_1$ and $J_2$ are the nearest and next nearest hopping strengths, and $V_1$ and $V_2$ are the nearest and the next nearest interaction strengths. Here we include the next nearest hopping and interaction to break the integrability. The SYK model contains random and all-to-all interactions, while the spinless fermion model has locality. These two models represent two different types of chaotic Hamiltonians. Aside from these two models, we have also numerically studied other models, such as the Ising model with transverse and longitudinal fields and the spin-$1/2$ Hubbard model. The results are similar. 

\textit{Behavior of $d_{nm}$.} In Fig. \ref{dnm}, we plot typical $d_{nm}$ for these two models \cite{code}. It is clear that the diagnoal matrix elements $d_{nn}$ (short-noted as $d_n$) are much larger than all off-diagonal matrix elements. For hermitian operators $\hat{M}_i$ and using the fact $\hat{W}^\dag_n=(-1)^n\hat{W}_n$, it is straightforward to show that \begin{equation}
d_n=\sum\limits_{i}\text{Tr}[ [\hat{W}_n, \hat{M}_i]^\dag [\hat{W}_n, \hat{M}_i]],  \label{dn}
\end{equation}
if plus sign is taken in Eq. \ref{dnm}. And 
\begin{equation}
d_n=\sum\limits_{i}\text{Tr}[ \{\hat{W}_n, \hat{M}_i\}^\dag \{\hat{W}_n, \hat{M}_i\}],  \label{dn-2}
\end{equation}
if minus sign is taken in Eq. \ref{dnm}, and anti-commutators replace commutators. In both cases, $d_n$ are always non-negative real numbers. 

\begin{figure}
\begin{center}
\includegraphics[width=0.45\textwidth]{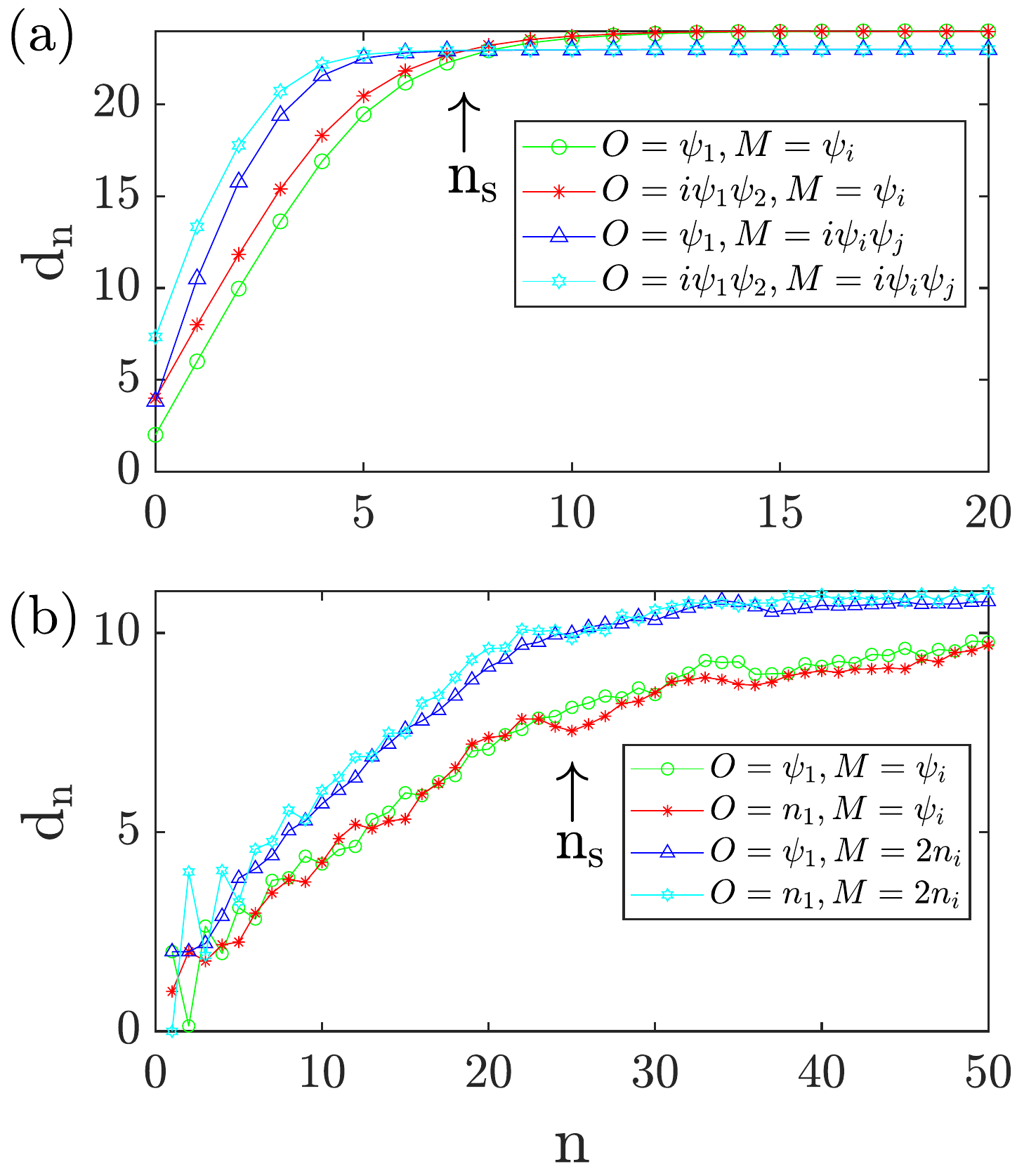}
\end{center}
\caption{Coefficients $d_{n}$ for the diagonal components of the non-hermitian terms for the SYK model (a) and for the spinless fermion model (b). For (a), four different curves cover four different cases: $\hat{O}$ and $\hat{M}_i$ are either fermionic operator $\hat{\psi}_i$ or bosonic operator $i\hat{\psi}_i\hat{\psi}_j$. We have set $J=1$. The system contains total number of Majorana fermion $N=24$.  For (b), four different curves cover four different cases: $\hat{O}$ and $\hat{M}_i$ are either fermionic operator $\hat{\psi}_i$ or bosonic operator $\hat{n}_i$, where $\hat{\psi}_i$ denotes $\hat{c}^\dag_i+\hat{c}_i$. We have set $J_1=1$, $J_2=0.2$, $V_1=0.6$ and $V_2=0.1$. The model contains total number of sites $N=13$. $n_\text{s}$ marks the places where $d_n$ saturates.  }
\label{fig:dn}
\end{figure}

Since Ref. \cite{Altman} has shown that $b_n$ increases linearly in $n$ for a generic chaotic Hamiltonian, here we focus on the behavior of $d_n$, as shown in Fig. \ref{fig:dn}. We find that $d_n$ also increases linearly with the increasing of $n$ and saturates when $n> n_\text{s}$. 

This behavior of $d_n$ can be understood as follows. Suppose $\hat{M}_i$ is a local operator at site-$i$, and if $\hat{W}_n$ acts trivially at site-$i$, then these two operators commute with each other and this commutator does not contribute to $d_n$. Hence, when operator size of $\hat{W}_n$ increases with the increasing of $n$, $d_n$ increases. Inspired by this argument, let us then assume $d_n$ increases as $\sim n^\delta$. Below we argue $\delta=1$. To this end, we ultilize the result of Ref. \cite{Altman} that in a closed system, Krylov complexity is a proper bound of the out-of-time-ordered commutator (OTOC). Here we consider the OTOC $\langle |[\hat{O}(t), \hat{M}_i]|^2\rangle$ of the closed system at infinite temperature, and we have 
\begin{align}
&\sum_i \langle |[\hat{O}(t),\hat{M}_i]|^2\rangle = \sum_i \langle |[\sum_n \varphi^\text{c}_n(t) \hat{W}_n,\hat{M}_i]|^2 \rangle\nonumber \\
& \approx \sum_i \sum_n |\varphi^\text{c}_n(t)|^2 \langle |[\hat{W}_n,\hat{M}_i]|^2\rangle= \sum_n |\varphi^\text{c}_n(t)|^2 d_n.
\end{align}
Here we use $\varphi^\text{c}_n(t)$ to denote the expansion coefficients of $\hat{O}(t)$ of the closed system. 
On the other hand, the infinite temperature OTOC $\langle |[\hat{O}(t), \hat{M}_i]|^2\rangle$ should be bound by $\mathcal{C}\mathcal{K}(t)$ in the same closed system, where $\mathcal{C}$ is certain constant. Hence, we have 
\begin{align}
 \sum_n d_n |\varphi^\text{c}_n(t)|^2  \leq\mathcal{C}\mathcal{K}(t)=\mathcal{C}\sum_n n|\varphi^\text{c}_n(t)|^2. \label{bound}
\end{align}
If this bound is valid, it requires $\delta<1$ and if the bound is tight, it further requires $\delta=1$. Hence, this gives rise to a linear growth of $d_n$. 

\begin{figure}
\begin{center}
\includegraphics[width=0.48\textwidth]{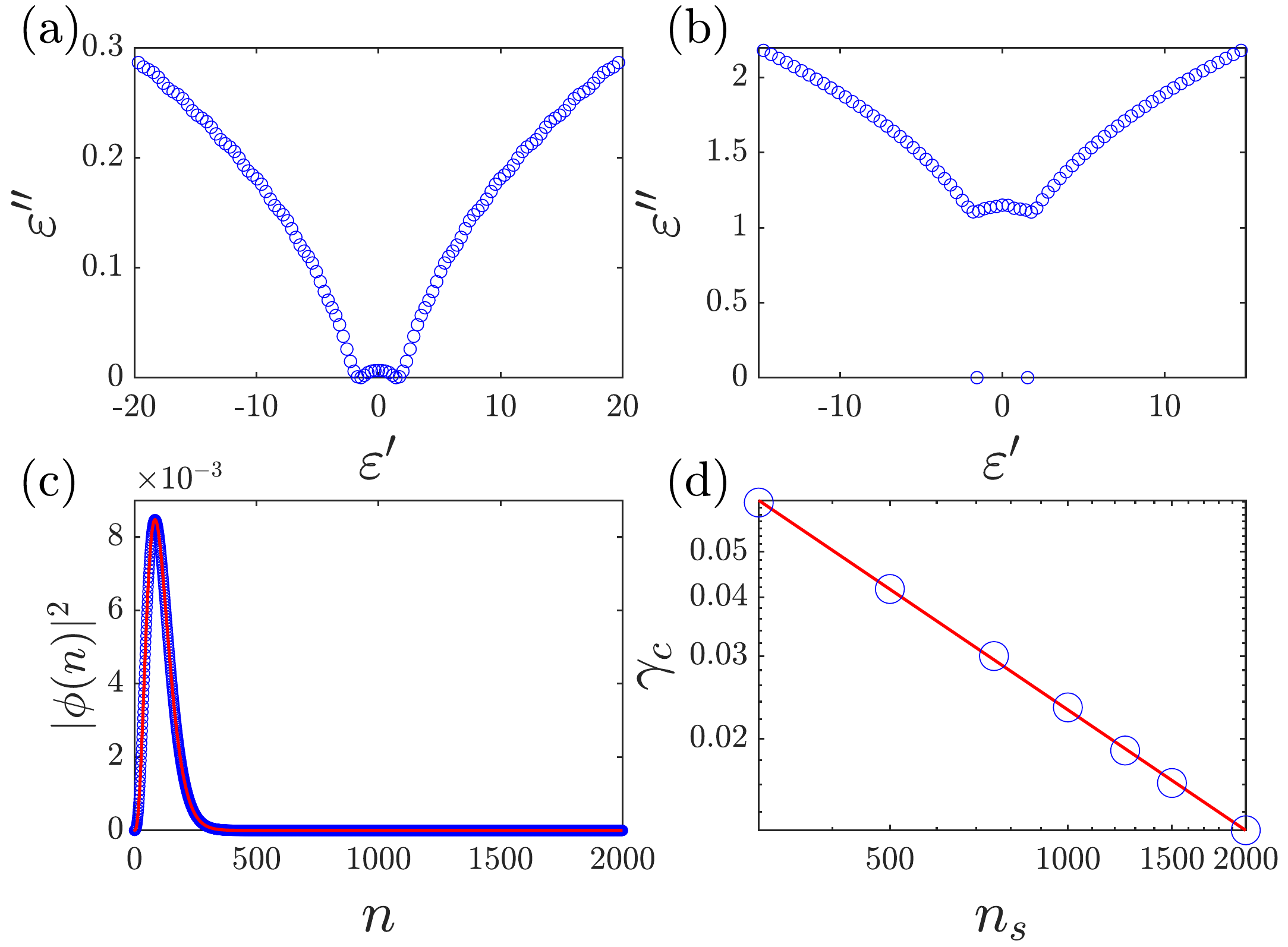}
\end{center}
\caption{Spectrum for the non-hermitian tight-binding model. (a-c) We choose $\beta_d=2.80$, $\alpha_d=0.35$, $\beta_b=0.66$, $\alpha_b=0.34$ and $n_\text{s}=1000$. (a) Eigen-energy $\epsilon=\epsilon^\prime-i\epsilon^{\prime\prime}$. $(\epsilon^\prime, \epsilon^{\prime\prime})$ for $\gamma=0.007<\gamma_\text{c}$ (b) $(\epsilon^\prime, \epsilon^{\prime\prime})$ for $\gamma=0.04>\gamma_\text{c}$. Here we shift the imaginary part $\epsilon^{\prime\prime}$ by a constant such that the smallest $\epsilon^{\prime\prime}$ is zero. (c) The wave functions for two modes below the gap in (b). (d) $\gamma_\text{c}$ versus $n_\text{s}$. }
\label{spectrum}
\end{figure}

Similar argument can be applied to the situation that both $\hat{O}$ and $\hat{M}$ are fermionic. In this case, the infinite temperature OTOC is also defined in terms of anti-commutator $\langle |\{\hat{O}(t), \hat{M}_i\}|^2\rangle$. Similarly, we have 
\begin{align}
&\sum_i \langle |\{\hat{O}(t),\hat{M}_i\}|^2\rangle = \sum_i \langle |\{\sum_n \varphi^\text{c}_n(t) \hat{W}_n,\hat{M}_i\}|^2\rangle\nonumber\\
& \approx \sum_i \sum_n |\varphi^\text{c}_n(t)|^2 \langle |\{\hat{W}_n,\hat{M}_i\}|^2\rangle= \sum_n |\varphi^\text{c}_n(t)|^2 d_n.
\end{align}
Also using Eq. \ref{bound}, we reach the same conclusion that $d_n$ increases linearly in $n$.

\textit{Spectrum of Non-hermitian Hopping Model.} Below we simplify the non-hermitian tight-binding model Eq. \ref{diff-non-hermitian} by only considering the diagonal components of $d_{nm}$. We assume that $b_n=\beta_b+\alpha_b n$ and $d_n=\beta_d+\alpha_d n$  up to $n=n_\text{s}$ and both remain constants for $n>n_\text{s}$, where $\beta_b$, $\alpha_b$, $\beta_d$ and $\alpha_d$ are always positive. We obtain reasonable values of these parameters by fitting Fig. \ref{fig:dn}(b).

The spectrum of the non-hermitian tight-binding model determines the dynamical behavior of $\mathcal{K}(t)$. We write the eigen-energies as $\epsilon=\epsilon^\prime-i\epsilon^{\prime\prime}$. The imaginary part $\epsilon^{\prime\prime}$ is always positive. We solve all the eigen-state $\phi_{l}$ of the non-hermitian tight-binding model and plot their eigen-energies $\{(\epsilon_l^\prime, \epsilon_l^{\prime\prime})\}$ in Fig. \ref{spectrum}.  We find that for $\gamma>\gamma_\text{c}$, the imaginary part of the spectrum acquires a gap $\sim\Delta$ as shown in Fig. \ref{spectrum}(b). Note that the time evolution in the half-infinite chain follows
\begin{equation}
\varphi(t)=\sum\limits_{l}c_l e^{-i(\epsilon^\prime_l-i\epsilon^{\prime\prime}_l)t}\phi_l.
\end{equation}
Hence, the eigen-modes with larger $\epsilon^{\prime\prime}$ decay faster. When $t\gg 1/\Delta$, all modes above the gap damps out and the mode below the gap dominates $\varphi(t)$. We show in Fig. \ref{spectrum}(c), the wave-functions of these two modes below the gap are localized around the edge of the half-infinite chain. We note that the existence of such localized edge states is a universal behavior of such non-hermitian tight-binding model. Therefore, $\mathcal{K}(t)$ saturates to a much smaller value at long-time, and this value is determined by the center position of the wave function shown in Fig. \ref{spectrum}(c). Such a behavior of $\mathcal{K}(t)$ is shown by the solid line in Fig. \ref{results}, compared with the closed system with the same Hamiltonian. This can also be understood from the fact that $d_n$ is larger for larger $n$, which means that more complicated operators are subjected to stronger decay, and therefore, simpler operators survive at long-time. Hence, we conclude that the growth of Krylov complexity is suppressed by dissipation. 

Here we also find that $\gamma_\text{c}$ depends on $n_\text{s}$. Fig. \ref{spectrum}(c) shows that $\gamma_\text{c}\sim 1/n^{0.85}_\text{s}$. $n_\text{s}$ increases as the total system size increases. Hence, for an infinite system, $n_\text{s}\rightarrow \infty$, and therefore, $\gamma_\text{c}\rightarrow 0$.

\textit{Summary and Outlook.} In summary, we generalize Krylov complexity to an open system governed by the Lindblad equation. We show that Krylov complexity defined for the open system can be mapped to a non-hermitian tight-binding model. This model also exhibits universal behavior, and its localized edge modes determine the long-time behavior of Krylov complexity in an open system. This work opens a new route to extend the discussion of operator complexity and chaos to open quantum systems. For those systems such as the SYK model with gravity interpretation, such discussion can also shed light on gravity physics through holographic duality. 

\textit{Acknowledgment.} We thank Pengfei Zhang for pointing out the minus sign in the operator Lindblad equation. We thank Zhenbin Yang for bringing our attention to Krylov complexity. We thank Wei Yi, Zhong Wang, Xiaoliang Qi, Yingfei Gu and Peng Zhang for helpful discussions. The project is supported by Beijing Outstanding Young Scholar Program, NSFC Grant No.~11734010 and the XPLORER Prize. 

\textit{Note Added.} When finishing this work, a new preprint appeared where Krylov complexity in an open system has also been proposed \cite{H. Sahu}. However, the method and the focus are different from ours.

\newpage

\begin{widetext}

\section{Supplementary Material}

In this supplementary material, we derive the Lindblad equation for an operator $\hat{O}(t)$. Especially, we should highlight that an extra minus sign is required in front of the quantum jump term when both $\hat{O}$ and the dissipation operator $\hat{M}$ are both fermionic ones. First of all, we consider the duality between the Schr\"odinger picture and the Heisenberg picture 
 \begin{equation}
 \text{Tr}[\hat{O}\hat{\rho}(t)]=\text{Tr}[\hat{O}(t)\hat{\rho}]. \label{O_def}
 \end{equation}
Since it is known that the density matrix $\hat{\rho}(t)$ of an open system obeys the following Lindblad equation \begin{equation}
\frac{d\hat{\rho}(t)}{dt}=-i[\hat{H},\hat{\rho}]+\gamma\sum\limits_{i}(2\hat{M}_i\hat{\rho}\hat{M}^\dag_i-\{\hat{M}^\dag_i\hat{M}_i,\hat{\rho}\}),
\end{equation}
it is straightforward to show that $\hat{O}(t)$ obeys the adjoint equation 
\begin{equation}
\frac{d\hat{O}(t)}{dt}=i[\hat{H},\hat{O}]+\gamma\sum\limits_{i}(2\hat{M}^\dag_i\hat{O}\hat{M}_i-\{\hat{M}^\dag_i\hat{M}_i,\hat{O}\}). \label{Lindblad_operator}
\end{equation} 
However, Eq. \ref{Lindblad_operator} is not always correct. When the operator $\hat{O}$ is fermionic, the expectation value of $\hat{O}(t)$ is always zero. As a result, Eq. \ref{O_def} cannot result in a unique equation for $\hat{O}(t)$ given the Lindblad equation for $\hat{\rho}(t)$. Therefore, in order to include the situations with both bosonic and fermionic operators $\hat{O}$, we shall derive the operator Lindblad equation by first explicitly including the operators in bath and then integrating out the bath operators.   

We start with the general form 
\begin{equation}
\hat{H} =  \hat{H}^\text{S}+\hat{H}^\text{B}+\hat{H}^{\text{int}},
\end{equation}
where $\hat{H}^\text{S}$ and $\hat{H}^\text{B}$ are respectively the Hamiltonians for system and for bath. $\hat{H}^{\text{int}}$ represents the interaction between system and bath, which is assumed to be 
\begin{equation}
\hat{H}^{\text{int}} = \lambda (\hat{M}^\dagger \hat\xi+\hat\xi^\dagger \hat{M}).
\end{equation}
The operator evolution in the entire system obeys the Heisenberg equation as
\begin{equation}
i\partial_t \hat{O} = [\hat{O},\hat{H}].
\end{equation}
The effective evolution of the operator acting on system only can be derived by tracing out the bath degree of freedom $\hat{O}^\text{S}(t) = \text{Tr}_\text{B}(\hat{\rho}_\text{B}\hat{O}(t))$. Below we turn into the interaction picture by introducing the unitary transformation $\hat{U}_0(t) = e^{-i (\hat{H}^\text{S}+\hat{H}^\text{B}) t}$. We write down the operator in the interaction picture as $\hat{O}_\text{I}(t) = \hat{U}_0(t)\hat{O}(t)\hat{U}_0^\dagger(t)$. Then, the evolution equation becomes
\begin{equation}
i\partial_t\hat{O}_\text{I} = [\hat{O}_\text{I},\hat{H}^{\text{int}}_{\text{I}}].  \label{OIt}
\end{equation}

We assume a Markovian bath and apply the white noise approximation to the Green's function of the bath, that is, 
\begin{equation}
\langle \hat\xi_\text{I}(t) \hat\xi_\text{I}^\dagger(t') \rangle = \delta(t-t'), \label{bath1}
\end{equation}
and we also have
\begin{equation}
\langle  \hat\xi_I(t) \rangle = \langle  \hat\xi_I^\dagger(t) \rangle = \langle  \hat\xi_I^\dagger(t) \hat\xi_I(t')\rangle = \langle  \hat\xi_I(t)\hat\xi_I(t') \rangle =  \langle  \hat\xi_I^\dagger(t)\hat\xi_I^\dagger(t') \rangle = 0. \label{bath2}
\end{equation}
Formally integrating out Eq. \ref{OIt} we obtain
\begin{equation}
\hat{O}_\text{I}(t) = \hat{O}_\text{I}(0) -i\int_0^t dt'[O_\text{I}(t'),H^{\text{int}}_\text{I}(t')].
\end{equation}
Substitute this equation back to Eq. \ref{OIt} we obtain
\begin{equation}
i\partial_t\hat{O}_\text{I}(t) = [\hat{O}_\text{I}(0),\hat{H}_\text{I}^{\text{int}}(t)]-i\int_0^t dt' [[\hat{O}_\text{I}(t'),\hat{H}_\text{I}^{\text{int}}(t')],\hat{H}_\text{I}^{\text{int}}(t)].
\end{equation}
Now we trace out the bath degree of freedoms, and it yields 
 \begin{align}
i\partial_t\hat{O}^S_\text{I}(t) &=\text{Tr}_\text{B}\left[ \hat\rho_\text{B}\left([\hat{O}^\text{S}_\text{I}(0),\hat{H}_\text{I}^{\text{int}}(t)]-i\int_0^t dt' [[\hat{O}^\text{S}_\text{I}(t),\hat{H}_I^{\text{int}}(t')],\hat{H}_\text{I}^{\text{int}}(t)] \right) \right ]
\label{partialO}\\
&= -i\int_0^t dt' \text{Tr}_\text{B} \left[\hat\rho_\text{B} (\hat{O}^\text{S}_\text{I}(t) \hat{H}^{\text{int}}_\text{I}(t') \hat{H}^{\text{int}}_\text{I}(t)+ \hat{H}^{\text{int}}_\text{I}(t) \hat{H}^{\text{int}}_\text{I}(t') \hat{O}^\text{S}_\text{I}(t) - \hat{H}^{\text{int}}_I(t') \hat{O}^\text{S}_\text{I}(t)\hat{H}^{\text{int}}_\text{I}(t)  - \hat{H}^{\text{int}}_\text{I}(t) \hat{O}^\text{S}_\text{I}(t) \hat{H}^{\text{int}}_\text{I}(t')  )\right]. \label{Ohh}
\end{align}
The first term in Eq.\ref{partialO} only contains single bath operator, therefore, thanks to Eq. \ref{bath2}, it vanishes after tracing out the bath. Since the kernal in the integration of Eq. \ref{Ohh} decays fast enough, we can extend the the upper limit of the integration to infinity. Then, using the bath correlation function Eq. \ref{bath1} and Eq. \ref{bath2}, we have
 \begin{equation}
 \begin{aligned}
i\partial_t\hat{O}^\text{S}_\text{I}(t) &=-i\lambda^2 \int_0^\infty dt' \text{Tr}_B[ \hat\rho_B (\hat{O}^\text{S}_\text{I}(t)  \hat{M}^\dagger_\text{I}(t') \hat{\xi}_\text{I}(t')    \hat{\xi}_\text{I}^\dagger(t) \hat{M}_\text{I}(t) +\hat{M}^\dagger_\text{I}(t) \hat{\xi}_\text{I}(t)   \hat{\xi}_\text{I}^\dagger(t') \hat{M}_\text{I}(t')   \hat{O}^\text{S}_\text{I}(t) \\
&\quad -  \hat{M}^\dagger_\text{I}(t) \hat{\xi}_\text{I}(t)   \hat{O}^\text{S}_\text{I}(t)    \hat{\xi}_\text{I}^\dagger(t') \hat{M}_\text{I}(t') 
-  \hat{M}^\dagger_\text{I}(t') \hat{\xi}_\text{I}(t')   \hat{O}^\text{S}_\text{I}(t)    \hat{\xi}_\text{I}^\dagger(t) \hat{M}_\text{I}(t) 
) ]\\
&= -i\lambda^2 \int_0^\infty dt' \langle\hat{\xi}_\text{I}(t)\hat\xi^\dagger_\text{I}(t')\rangle (\{\hat{M}_\text{I}^\dagger(t)\hat{M}_\text{I}(t),\hat{O}^\text{S}_\text{I}(t)\}
-2(-1)^\eta \hat{M}_\text{I}^\dagger(t)\hat{O}^\text{S}_\text{I}(t) \hat{M}_\text{I}(t))\\
&= -i\lambda^2  (\{\hat{M}_\text{I}^\dagger(t)\hat{M}_\text{I}(t),\hat{O}^\text{S}_\text{I}(t)\}
-2(-1)^\eta \hat{M}_\text{I}^\dagger(t)\hat{O}^\text{S}_\text{I}(t) \hat{M}_\text{I}(t)),
\end{aligned}
\label{tricky}
\end{equation}
where the index $\eta$ comes from exchanging operator $\hat\xi$ or $\hat\xi^\dagger$ with operator $\hat{O}^\text{S}_\text{I}$ in system. When $\hat{M}$ is fermionic, $\hat{\xi}$ should also be fermionic. Then, if $\hat{O}$ is also fermionic, there will be an extra minus sign when exchanging $\hat{\xi}$ or $\hat{\xi}^\dag$ with $\hat{O}$, and therefore, $\eta = 1$. Otherwise, if either $\hat{\xi}$ or $\hat{O}$ is bosonic, or both of them are bosonic, there will be no extra sign when exchanging $\hat{\xi}$ or $\hat{\xi}^\dag$ with $\hat{O}$, and therefore, $\eta = 0$. Then, we return to the Heisenberg picture from interaction picture by the unitary transformation $\hat{O}^S(t) = \hat{U}^\dagger_0(t)\hat{O}^S_\text{I}(t)\hat{U}_0(t)$. Hence, we finally reach the operator Lindblad equation 
\begin{equation}
\frac{d\hat{O}^S(t)}{dt}=i[\hat{H},\hat{O}^S]-\gamma \left(\{\hat{M}^\dag\hat{M},\hat{O}^S\}-(-1)^\eta2\hat{M}^\dag\hat{O}^S\hat{M}\right). 
\end{equation} 
where $\gamma = \lambda^2$. When $\eta=0$, it is consistent with Eq. \ref{Lindblad_operator}.

There is a tricky point that worths mentioning in Eq.\ref{tricky}. Considering the term $\hat{M}^\dagger_\text{I}(t) \hat{\xi}_\text{I}(t)   \hat{O}^\text{S}_\text{I}(t)    \hat{\xi}_\text{I}^\dagger(t') \hat{M}_\text{I}(t') $, we can also write it as  
\begin{equation}
\text{Tr}_B[\hat{\rho}_{B}\hat{M}^\dagger_\text{I}(t) \hat{\xi}_\text{I}(t)   \hat{O}^\text{S}_\text{I}(t)    \hat{\xi}_\text{I}^\dagger(t') \hat{M}_\text{I}(t') ]=\text{Tr}_B[\hat{\rho}_{B}\hat{\xi}_\text{I}(t) \hat{M}^\dagger_\text{I}(t)    \hat{O}^\text{S}_\text{I}(t)     \hat{M}_\text{I}(t')\hat{\xi}_\text{I}^\dagger(t') ]
\end{equation}  
In order to trace out the bath operators, we need to move $\hat{\rho}_{\text{B}}$, $\hat{\xi}$ and $\hat{\xi}^{\dagger}$ together. There are seemingly two different methods to do so. The first method is to utilize the partial-trace's permutation as follows:
\begin{align}
\text{Tr}_B[\hat{\rho}_{B}\hat{\xi}_\text{I}(t) \hat{M}^\dagger_\text{I}(t)    \hat{O}^\text{S}_\text{I}(t)     \hat{M}_\text{I}(t')\hat{\xi}_\text{I}^\dagger(t') ]&=\text{Tr}_B[\hat{\xi}_\text{I}^\dagger(t') \hat{\rho}_{B}\hat{\xi}_\text{I}(t) \hat{M}^\dagger_\text{I}(t)    \hat{O}^\text{S}_\text{I}(t) \hat{M}_\text{I}(t')]\nonumber\\
&=\text{Tr}_B[\hat{\xi}_\text{I}^\dagger(t') \hat{\rho}_{B}\hat{\xi}_\text{I}(t)]\hat{M}^\dagger_\text{I}(t)    \hat{O}^\text{S}_\text{I}(t) \hat{M}_\text{I}(t')\nonumber\\
&=\text{Tr}_B[ \hat{\rho}_{B}\hat{\xi}_\text{I}(t)\hat{\xi}_\text{I}^\dagger(t')]\hat{M}^\dagger_\text{I}(t)    \hat{O}^\text{S}_\text{I}(t) \hat{M}_\text{I}(t').
\label{tricky1}
\end{align}
The second method is to directly pass $\hat{\xi}^{\dagger}$ through $\hat{M}^{\dagger} \hat{O}^S \hat{M}$ 
\begin{align}
\text{Tr}_B[\hat{\rho}_{B}\hat{\xi}_\text{I}(t) \hat{M}^\dagger_\text{I}(t)    \hat{O}^\text{S}_\text{I}(t)     \hat{M}_\text{I}(t')\hat{\xi}_\text{I}^\dagger(t') ]&=(-1)^{\eta}\text{Tr}_B[\hat{\rho}_{B}\hat{\xi}_\text{I}(t)\hat{\xi}_\text{I}^\dagger(t')  \hat{M}^\dagger_\text{I}(t)    \hat{O}^\text{S}_\text{I}(t)     \hat{M}_\text{I}(t')]\nonumber\\
&=(-1)^{\eta}\text{Tr}_B[\hat{\rho}_{B}\hat{\xi}_\text{I}(t)\hat{\xi}_\text{I}^\dagger(t')]  \hat{M}^\dagger_\text{I}(t)    \hat{O}^\text{S}_\text{I}(t)     \hat{M}_\text{I}(t').\label{tricky2}
\end{align}
This method generates a minus sign in the case when $\hat{M}^{\dagger}$, $\hat{O}^{\text{S}}$, $\hat{M}$ and $\hat{\xi}^{\dagger}$ are all fermionic. Obviously, these two methods contradict with each other once this minus sign is present. Here we should argue that the first method is not correct when $\hat{\xi},\hat{\xi}^{\dagger}$ are fermionic. 

Let us consider the total Hilbert space $\mathcal{H}_{\text{total}}$ as a tensor product of system and bath $\mathcal{H}_\text{S}\otimes\mathcal{H}_\text{B}$. And let us consider $\hat{M}^{\dagger}$, $\hat{M}$ and $\hat{O}^{\text{S}}$ are supported solely on $\mathcal{H}_\text{S}$. When $\hat{\xi}$ and $\hat{\xi}^{\dagger}$ are bosonic operators, their support are solely on $\mathcal{H}_\text{B}$. Then, $\hat{\xi}^{\dagger}$ and $\hat{\xi}$ commute with $\hat{M}^{\dagger}\hat{O}^{S}\hat{M}$. Both methods are consistent with each other. However, when $\hat{\xi}$ and $\hat{\xi}^{\dagger}$ are fermionic, although they are bath operators, they are supported on the total Hilbert space $\mathcal{H}_{\text{total}}$ in order to fulfil the anti-commutation relation with fermion operators in $\mathcal{H}_\text{S}$. In other word, the matrix representation of $\hat{\xi}$ or $\hat{\xi}^\dag$ requires a sign that depends on the physical state of the system. This is reminiscent of the Jordan-Wigner transformation where fermions carry non-local strings in its matrix representation. Thus, the partial-trace's permutation, i.e. the first equality in Eq. \ref{tricky1}, is not correct. This difference between bosonic and fermionic operators is explicitly illustrated in Fig. \ref{FIG_tricky}. Moreover, in fermionic case, one can alternatively assume that $\hat{\xi}$ and $\hat{\xi}^\dag$ are solely supported on $\mathcal{H}_\text{B}$. Then, $\hat{M}^{\dagger}$, $\hat{M}$ and $\hat{O}^{\text{S}}$ should supported on the entire $\mathcal{H}_{\text{total}}$ in order to fulfill the anti-commutation relation. Then, the second equality in Eq. \ref{tricky1} is not correct. In any case, the first method fails. The same discussion applies to another term $\hat{M}^\dagger_\text{I}(t') \hat{\xi}_\text{I}(t')   \hat{O}^\text{S}_\text{I}(t)    \hat{\xi}_\text{I}^\dagger(t) \hat{M}_\text{I}(t)$ in Eq. \ref{tricky}.

\begin{figure}[t]
\begin{center}
\includegraphics[width=1\textwidth]{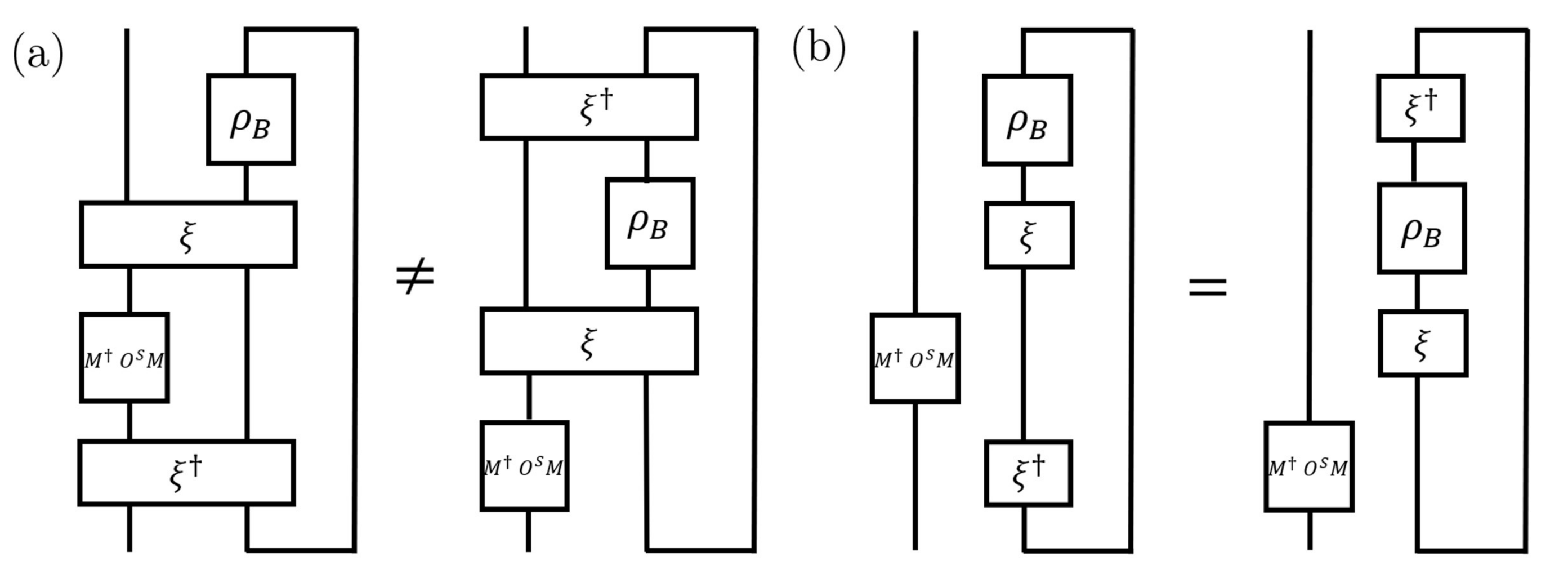}
\end{center}
\caption{A quantum circuit illustration of evaluation of $\text{Tr}_B[\hat{\rho}_{B}\hat{\xi}_\text{I}(t) \hat{M}^\dagger_\text{I}(t)    \hat{O}^\text{S}_\text{I}(t)  \hat{M}_\text{I}(t')\hat{\xi}_\text{I}^\dagger(t') ]$. (a) When $\hat{\xi}$ and $\hat{\xi}^{\dagger}$ are fermionic, they are supported on the total Hilbert space and the partial-trace's permutation is incorrect. (b) When $\hat{\xi}$ and $\hat{\xi}^{\dagger}$ are bosonic, they are only supported on the Hilbert space of bath and the partial-trace's permutation is correct.}
\label{FIG_tricky}
\end{figure}

\end{widetext}

\end{document}